\documentclass[11pt,a4paper]{article}

\usepackage{graphicx}
\usepackage{amssymb}

\usepackage{times}
\newcommand{\bea}{\begin{eqnarray}}  
\newcommand{\eea}{\end{eqnarray}}

\usepackage{graphicx}
\usepackage{amssymb}

\begin{document}
\begin{Large}
\begin{center}
\textbf{Turning with the others: novel transitions in an SPP model with coupling of accelerations}
\end{center}
\end{Large}

\medskip

\begin{large}
\begin{center}
P\'eter Szab\'o$^{\dagger\nmid}$,
M\'at\'e Nagy$^{\dagger}$, and
Tam\'as Vicsek$^{\dagger\ddagger}$
\end{center}
\end{large}

\bigskip 
 
$^{\dagger}$Department of Biological Physics, E\"otv\"os University,  
P\'azm\'any P\'eter s\'et\'any 1A, H-1117 Budapest, Hungary 
$^{\ddagger}$Statistical and Biological Physics Research Group  
of the Hungarian Academy of Sciences, P\'azm\'any P\'eter s\'et\'any 1A, 
H-1117 Budapest, Hungary 
$^{\nmid}$To whom correspondence should be addressed.  
E-mail: pszabo@angel.elte.hu. 
    
\bigskip



\begin{abstract}
We consider a three dimensional, generalized version of the
original SPP model  for collective motion.
By extending the factors influencing the ordering, we investigate
the case when the movement of the self-propelled particles (SPP-s)
depends on both the velocity and the acceleration of the
neighboring particles, instead of being determined solely by the
former one. By changing the value of a weight parameter $s$
determining the relative influence of the velocity and the
acceleration terms, the system undergoes a kinetic phase
transition as a function of a behavioral pattern.
Below a critical value of $s$ the system exhibits
disordered motion, while above it the dynamics resembles that of
the SPP model. We argue that in nature evolutionary processes can
drive the strategy variable $s$ towards the critical point, where
information exchange between the units of a system is maximal.
\end{abstract}


\section{Inroduction}

Collective motion of organisms (e.g. fish schools, bird flocks, 
bacterial colonies) exhibits a large variety of
emergent phenomena \cite{Inagaki_et_al_1976, Feare_book_1984, Ben-Jacob_et_al_1995, Rauch_et_al_1995, Shapiro_1995, Parrish_1999, Szabo_et_al_2006, Wu_2006}.
Synchronized motion, symmetrical group formations
(e.g., V shaped) or swirling patterns emerge in spite of the
apparently simple behavioral rules of the individual flock members
\cite{Parrish_Hammer_book_1997, Couzin_and_Krause_2003}.
The self propelled particles (SPP) model was proposed by Vicsek
{\it et al.} \cite{Vicsek_et_al_1995} to describe the
onset of ordered motion within a group of self-propelled particles
in the presence of perturbations. Taking into the effects of
fluctuations inevitably present in biological systems was an
essential generalization of the prior deterministic flocking
models such as that of Reynolds \cite{Reynolds_1987}.
The original model considers point-like particles
moving at constant velocity on a two dimensional surface with
periodic boundary conditions. The only rule is that, at each time
step, every particle approximates, with some uncertainty, the average
direction of motion of the particles within its neighborhood of
radius $R$. This model exhibits spontaneous self-organization; by
decreasing the noise parameter, the system undergoes a kinetic
phase transition from a disordered state to an ordered one
where all the particles move approximately in the same direction.
Due its simplicity and analogy with biological systems
comprised of many, locally interacting particles, the SPP model
soon became a reference model for the flocking behavior of
organisms \cite{Czirok_et_al_1997, Csahok_and_Vicsek_1995,Toner_and_Tu_1995,
Toner_and_Tu_1998,Toner_et_al_1998,Czirok_et_al_1999,
Gregoire_and_Chate_2001,Gregoire_and_Chate_2004,Nagy_et_al_2007,Huepe_2008, Chate_et_al_2008}.

The individual based behavioral rules, determining collective
motion, are of particular interest. Important elements of behavioral
rules are the nature of the perceived information and the affected
behavioral traits.
\cite{Ballerini_et_al_2008,Theraulaz_et_al_2008a,Theraulaz_et_al_2008b}.
A frequent assumption in models is that the information,
perceived by the particles, is restricted to the velocity of their
neighbors. The interaction range is usually defined by metric
distances, but Ballerini {\it et al.} \cite{Ballerini_et_al_2008}
recently showed that topological distance is the one determining the
flocking of starlings.
The assumption of reflecting on the momentary velocity only may not be
enough for adequately describing a number of biologically relevant 
situations. We expect that the behavior of the SPP model is
significantly extended if we also incorporate a term corresponding
to memory on short time scales. This can be achieved by
introducing an acceleration term into the equations (rules).
This is equivalent to separating time scales by assuming that the 
particles differentiate between two kind of information, the first
being their actual velocity, the second one corresponding to recent change
in their direction of motion. For example, in the case of birds, reacting
to acceleration may mean that birds can give signals to their neigbors about
their intended changing of their flight direction by quickly modifying
their velocity.

\section{Model}

In the three-dimensional, scalar noise (SNM) version \cite{Gonci_et_al_2008}
of the original SPP model \cite{Vicsek_et_al_1995} the particles are
assumed to move with a constant velocity $\nu$, and their
positions are updated simultaneously according to
\begin{equation}
\mathbf{x_i}(t+\Delta t)=\mathbf{x_i}(t)+\mathbf{v_i}(t)\Delta t,
\label{eq_snm_pos}
\end{equation}
where $\mathbf{x}_i$ and $\mathbf{v}_i$ are position and velocity
of particle $i$, respectively. The time increment is set to be $\Delta t=1$.
Each particle is assumed to move, with some uncertainty, in the average direction of all
the particles within a fixed neighborhood of radius $R=1$. Hence, the new velocity is given by
\begin{equation}
\mathbf{v_i}(t+\Delta t)=\nu \cdot \mathcal{M}(\mathbf{e},\xi) \cdot
\mathbf{N} \left( \langle \mathbf{v}(t) \rangle_{i,R} \right),
\label{eq_snm_vel}
\end{equation}
where $\nu$ is the absolute value of velocity,
$\mathcal{M}(\mathbf{e},\xi)$ is a rotational tensor
representing a random perturbation, $\langle
\mathbf{v}(t) \rangle_{i,R}$ denotes the average velocity of all
particles around particle $i$ within radius $R$ including particle
$i$ itself, and $\mathbf{N(u)=u/|u|}$.
$\mathcal{M}(\mathbf{e},\xi)$ performs a rotation of angle
$\xi$ around a vector $\mathbf{e}$; $\xi$ is a uniform random
value in the interval $[-\eta \pi,\eta \pi]$, whereas
$\mathbf{e}$ is a random unit vector chosen uniformly from the set
of vectors perpendicular to $\mathbf{N} \left( \langle
\mathbf{v}(t) \rangle_{i,R} \right)$. 
The order in which (\ref{eq_snm_pos}) and (\ref{eq_snm_vel}) are
calculated has some quantitative effects on the results (see later).

Here we introduce the acceleration coupled self-propelled particles model (AC-SPP)
being a modified version of SNM, in which the velocity vector
$\mathbf{v_i}(t+\Delta t)$ is a function of both the velocity
$\mathbf{v}(t)$ and the acceleration
$\mathbf{a}(t)=(\mathbf{v}(t)-\mathbf{v}(t-\Delta t))/\Delta t$ of
the neighboring particles. Then equation \ref{eq_snm_vel} becomes

\begin{equation}
\mathbf{v_i}(t+\Delta t) = \nu \cdot
\mathcal{M}(\mathbf{e},\xi) \cdot
\mathbf{N}
\left(
s \cdot \langle \mathbf{v}(t) \rangle_{i,R} +
(1-s) \cdot \langle \mathbf{a}(t) \Delta t \rangle_{i,R}
\right),
\end{equation}
where $s \in (0,1]$ is a so-called strategy parameter, expressing
the relative influence of the acceleration and velocity tags on
the velocity vector of the focal particle. Initially, we have
$N=\rho L^3$ randomly distributed particles, where $L$
and $\rho$ stand for box size and particle number density,
respectively. The bounding box has periodic boundary conditions.
The velocity parameter used in the simulations is $\nu=0.1$,
corresponding to the low velocity regime \cite{Nagy_et_al_2007}.

Taking into account the acceleration in a separate term has
various possible motivations. As for the contribution of the
$i$-th particle, it corresponds to a memory effect: a given
particle, if it has no neighbors, has a tendency to keep on turning
as it did in the previous time step. Perhaps more importantly,
the average turning rate of the neighbors has now a separated
effect on the turning rate of the $i$-th particle. In the $s$
close to 1 limit the AC-SPP model is very much like the original
SPP, while for $s \ll 1$ the acceleration term dominates and
instantaneous turning of the neighbors has a strong effect on the
trajectory of the $i$-th particle. In this way not only the state
(velocity), but a kind of behavioral pattern (turning) can be
taken into account. We associate, for example, with such a
behavioral pattern a short turning period of a bird flying in a
flock and giving sign for its neighbors of its intended changing
of the direction of flight.

We characterize the collective motion of particles by the average velocity
of all particles $\varphi$, defined as
\begin{displaymath}
\varphi=\frac{1}{N} \Bigg| \sum_{i=1}^{N} \mathbf{v_i} \Bigg|,
\end{displaymath}
with $N$ denoting the number of particles in the system. This
order parameter can take any value in the range $[0,1]$, and
expresses the tendency of particles to move in the same direction.
If the particles move randomly $\varphi=0$ whereas if every
particle moves in the same direction $\varphi=1$. $\langle \varphi
\rangle$ was obtained by averaging over $12$ individuals runs,
each recorded after a relaxation time to stationarity $T_{relax}$ and with
time averaging over an additional $T_{avr}$ time steps.

\section{Results}

\begin{figure}[h!tbp]
{
\centerline{\includegraphics[width=0.5\columnwidth]{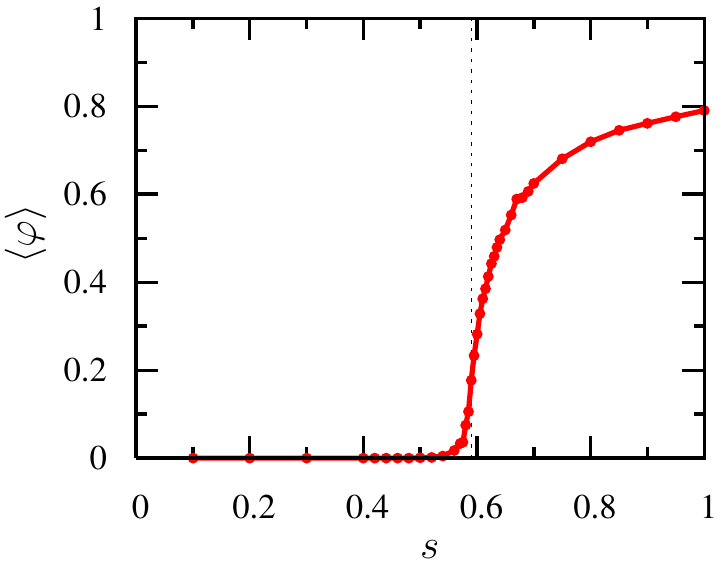}}}
\caption{Average velocity $\varphi$ as a function of the strategy variable $s$.
The data points were obtained by averaging the results of $12$
simulations, using the parameters $L=100$, $\rho=0.16$, $\nu=1/9$,
with a relaxation time of $T_{relax}=10000$ time steps and
averaging over an additional $T_{avr}=10000$ time steps.
\label{avr_vel}}
\end{figure}

At first we investigated the dynamics of the system at different
values of the strategy variable $s$ at fixed density and noise
values: $\rho=0.16$, $\eta=1/9$. The density was choosen to be high
enough to get flocking, but allowing moderate CPU times per run.
The noise value was a typical value resulting in an ordered state
for the original SPP model ($s=1$).
We found that by increasing $s$
the system undergoes a phase transition; below a critical value,
$s_c$, the ordering, expressed by $\varphi$, is negligible, while
above $s_c$ the level of order increases rapidly as a function of
$s$ (Figure \ref{avr_vel}). This is a novel type of phase
transition since it corresponds to a phase change {\it due to a change
in the relative strength of a behavioral pattern}.

\begin{figure}[h!tbp]
{
\centerline{\includegraphics[width=0.5\columnwidth]{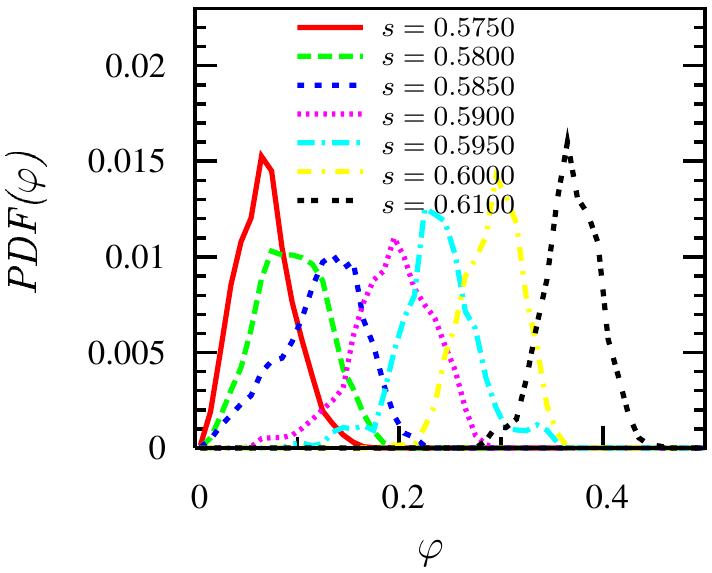}}}
\caption{Probability density function (PDF) of the order parameter
$\varphi$. Parameters as in the previous plot.
The one-humped distributions in a range of $s$ values
close to the critical one ($s_c \simeq 0.59$) suggest a second order
phase transition. 
\label{velpdf}}
\end{figure}

\begin{figure}[h!tbp]
{
\centerline{\includegraphics[width=0.5\columnwidth]{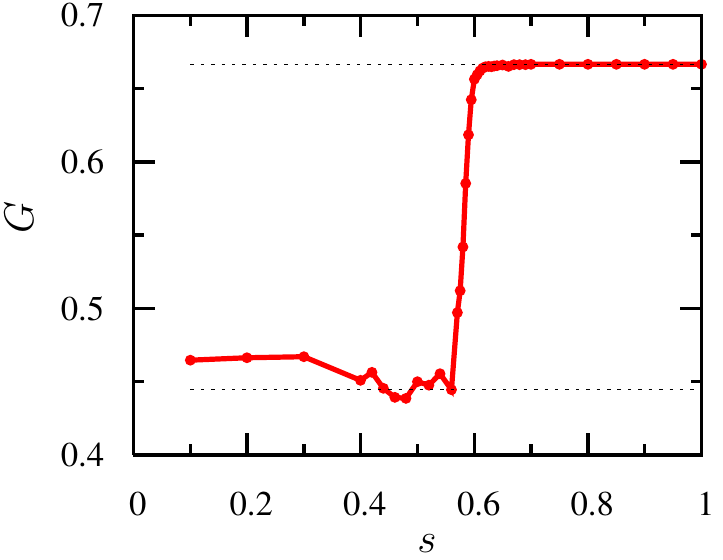}}}
\caption{The Binder cumulant $G$ as a function of the strategy variable $s$.
The dotted lines at 2/3 and 4/9 indicate the theoretical value in the 
case of the ordered and totally disordered states, respectively.
Parameters as in the previous plots. The absence of a sharp minimum
indicates a second order phase transition.
\label{binder}}
\end{figure}

In order to determine the nature of the phase transition we also
calculated the probability density function (PDF) of $\varphi$ and
the Binder cumulant $G$ at different strategy values. The Binder
cumulant, defined as $G=1-\langle \varphi^4 \rangle /3 \langle
\varphi^2 \rangle^2$, measures the fluctuations of the order
parameter and can be used to distinguish between first- and second
order phase transitions \cite{Binder_and_Herrmann_book_1997}.
In case of a first order phase transition $G$ exhibits a
characteristic minimum, whereas in case of a second order
transition this sharp minimum is absent. In our case, the
PDF was unimodal, and $G$ did not have a sharp minimum around
$s_c$, both indicating a second order phase transition (Figures
\ref{velpdf} and \ref{binder}). $G$ was monotonously increasing,
as it was observed already in a three dimensional SNM with 
continuous phase transition \cite{Gonci_et_al_2008}.

\begin{figure}[h!tbp]
{
\centerline{\includegraphics[width=0.5\columnwidth]{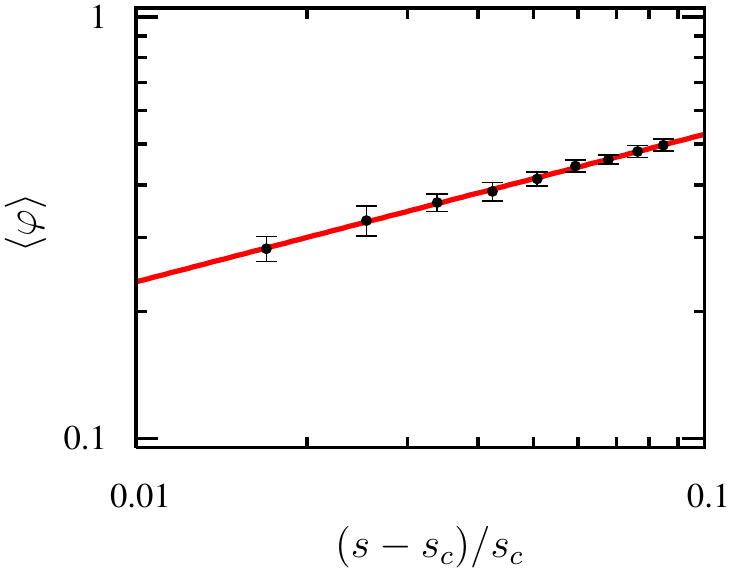}}}
\caption{Average velocity values as a function of the distance from the critical point.
The fitted line, with $s_c=0.590 \pm 0.002$, had a slope of $\beta=0.34 \pm 0.05$.
Parameters as in the previous plots.
\label{criticalvalues}}
\end{figure}

Consequently, near the critical point, the order parameter obeys the scaling relation
\begin{equation}
\varphi \sim \left[s - s_c(\rho,\eta) \right]^\beta,
\end{equation}
where $\beta$ is the critical exponent. The critical values $s_c$
and $\beta$ were determined by plotting $log \langle \varphi
\rangle$ as a function of $log[(s-s_c)/s_c]$ (Figure
\ref{criticalvalues}). $s_c$ was obtained by finding the value
where the plot was the straightest in the relevant region, whereas
the critical exponent is equal to the slope of the fitted line. 
We obtained $s_c=0.590 \pm 0.002$ and $\beta=0.35 \pm 0.05$.
It should be noted, that the value $\beta=0.35$ is
far from 0.5 (corresponding to mean-field or bifurcation)
indicating the true non-trivial nature of the transition we
describe.

\begin{figure}
{
\centerline{\includegraphics[width=0.5\columnwidth]{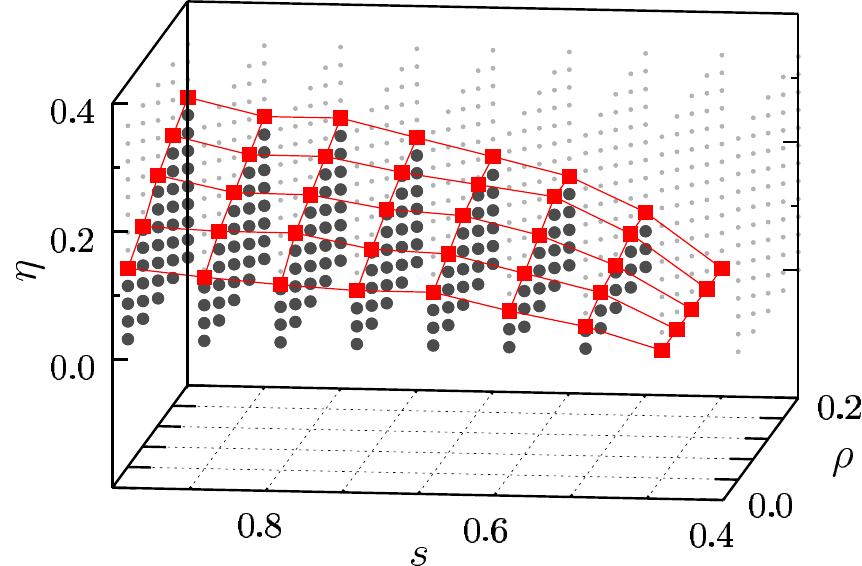}}}
\caption{Border surface between the ordered and disordered phases in the $s$, $\rho$, $\eta$
parameter space. The ordered ($\langle \varphi \rangle \ge 0.01$) and
disordered ($\langle \varphi \rangle < 0.01$) states are indicated by heavy
and light dots, respectively. Squares show the critical noise values $\eta_c(s,\rho)$
obtained from the plot of $log \langle \varphi \rangle$ as a function
of $log[(\eta_c-\eta)/\eta_c]$ for each $s$ and $\rho$ pair. $L=100$.
\label{paramspace}}
\end{figure}

The critical value of the transition depends on both the density and the noise parameters.
Figure \ref{paramspace} shows that $s_c$ is decreasing with both
increasing density and noise values.

The order variable $\varphi$, by itself, provides a poor
description of the two states, hence we also calculated other
statistics. The sinuosity of the particle
trajectories was expressed by their average curvature, defined as
$\lambda=\frac{1}{N}\sum_N \frac{|v_i \times a_i|}{|v_i|^3}$.
Two other statistics, $\psi$ and $\mu$, measure the information exchange
between the particles. $\psi$ is defined as the average number of
different particles encountered, i.e., being within a distance
$R$, by a focal particle during a given time interval.
Although this quantity depends on the chosen time interval,
its monotonicity as a function of $s$ is independent of it.
$\mu$ was used to evaluate the speed of information propagation, as follows.
Initially one percent of the particles held the information.
The information was transmitted between particles via encounters
between information holders and other particles.
This way sooner or later every particle became information holder.
$\mu$ was defined as the time needed for at least 90 percent of
the particles becoming information holders. 

All the three statistics $\langle \lambda \rangle$,
$\langle \psi \rangle$ and $\langle \mu \rangle$ were obtained by
averaging over $12$ individual runs, each with a relaxation
time of $T_{relax}=10000$ and averaging time of $T_{avr}=10000$.
$\langle \psi \rangle $ is an average value for 100 randomly
chosen particles.

\begin{figure}[h!tbp]
{
\centerline{\includegraphics[width=0.5\columnwidth]{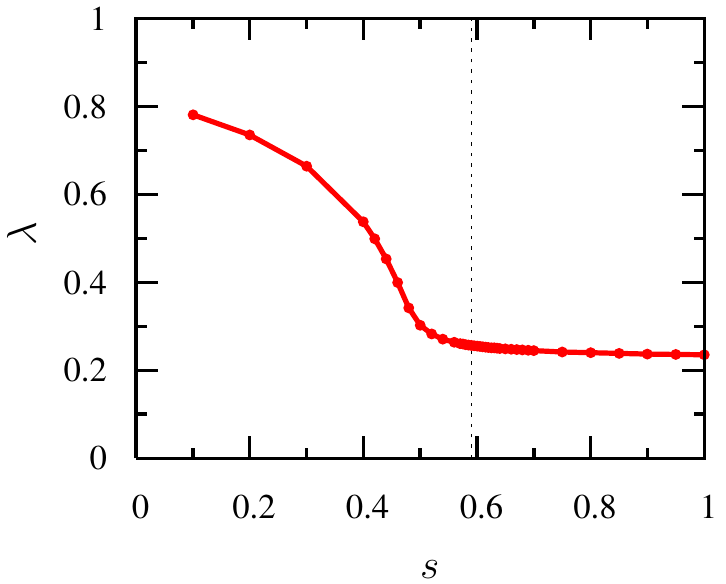}}}
\caption{The curvature of trajectories $\lambda$ as a function of the strategy variable $s$.
The dotted line at $s_c=0.590$ indicates the position of the critical point for ordering.
Parameters as in the previous plots.
\label{curvature}}
\end{figure}

\begin{figure}[h!tbp]
{
\centerline{\includegraphics[width=1.0\columnwidth]{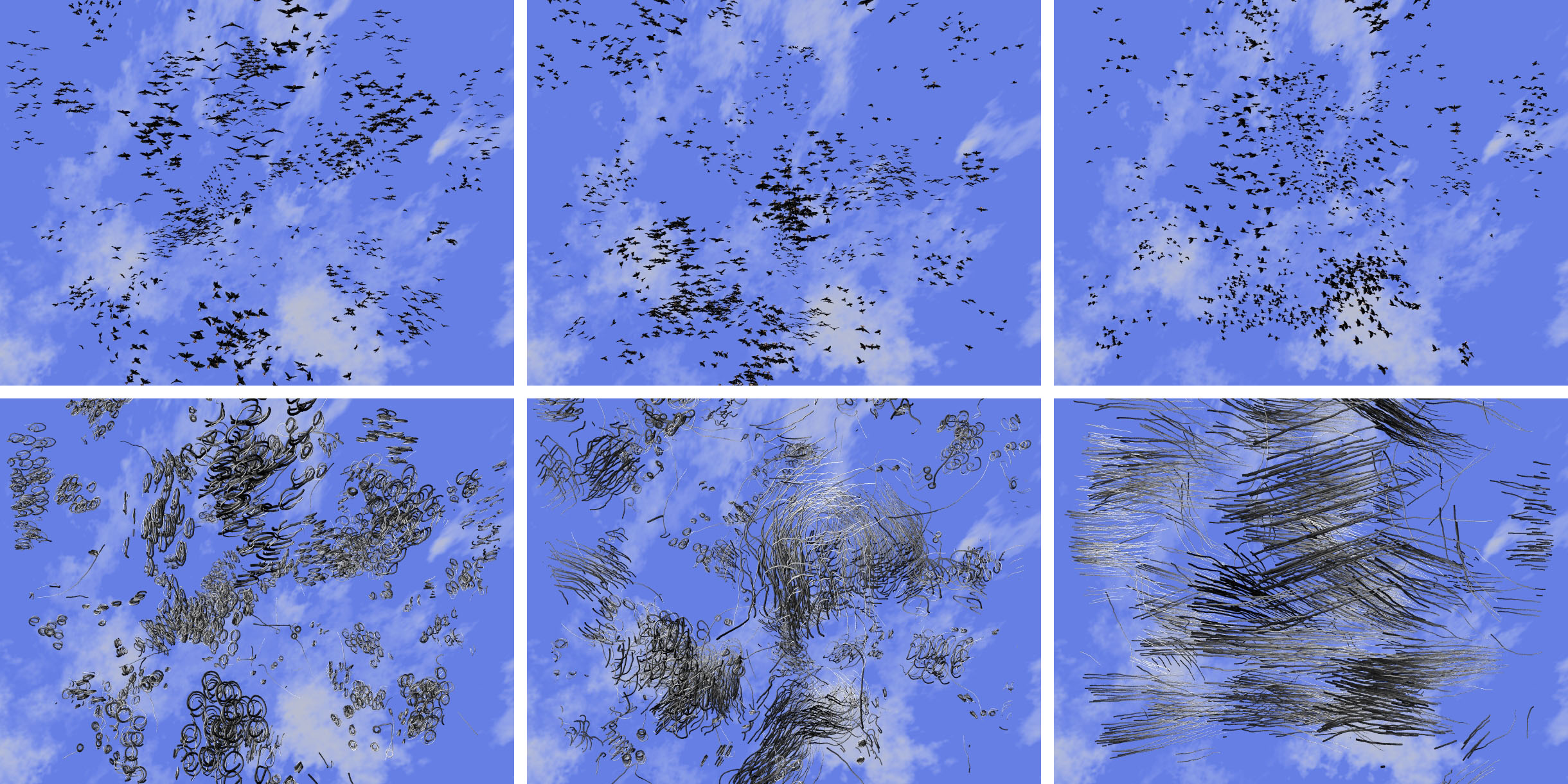}}}
\caption{Positional data and spatial trajectories of particles at different strategy values.
Subfigures show the typical behavior (a, d) below, (b, e) near and (c, f) above the critical point, 
respectively. (a-c) Positional data of the particles shown as birds.
(d-f) Each curve shows the trajectory of a particle over 60 time steps
after reaching steady state ($t=20000$). Different shades of gray indicate the time past, with darker tones denoting more recent positions.
(a, d) Below the critical point, at $s=0.5$ the particles move in circles.
(b, e) Near the critical point, at $s=0.53$, the particles move sinuously.
(c, f) Above the critical point, at $s=0.9$, the particles move in the same direction.
$L=20$, $\eta=1/6$, other parameters as in the previous plots.
\label{snapshots}}
\end{figure}

The curvature of the trajectories decreases with the strategy variable
(Figure \ref{curvature}). Large curvature at low $s$ values
indicates in this case, that particles move in small circles
(Figure \ref{snapshots}). It is because the large influence of the
acceleration term results in continuous turning. This turning is
synchronized among neighbors, i.e., their acceleration and
velocity vectors become the same, resulting in a particle cloud
consisting of separated groups of particles, each containing
circling particles. By increasing $s$ the radius of these circles
increases, until the circling groups overlap and start to interact
with each other. At a critical point the circles overlap so much,
that neither their position, nor their composition remains the
same; in other words the circling groups lose their identity and
the particles start to move sinuously. At large $s$ values the
movement becomes ordered; all particles tend to move in the same
direction, similar to the ordered phase of the SNM model at small
velocities.

\begin{figure}[h!tbp]
{
\centerline{\includegraphics[width=0.5\columnwidth]{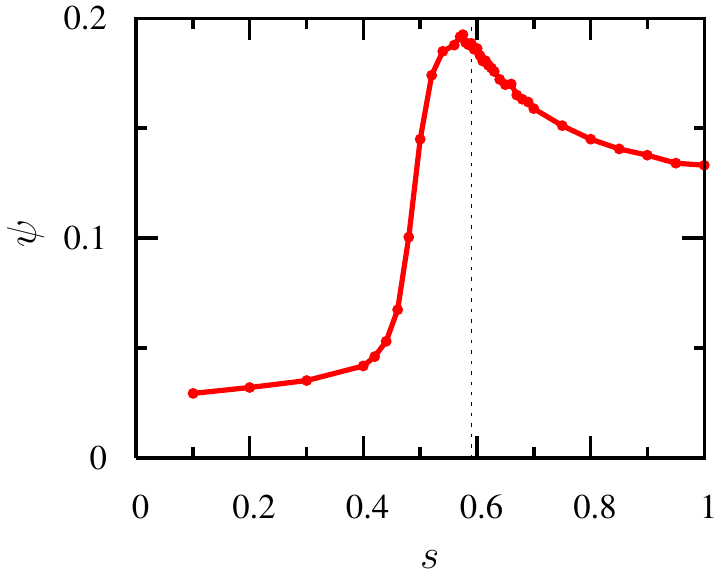}}}
\caption{Information exchange between particles ($\psi$) as a function of the strategy variable $s$. The curve for $\mu$ was very similar (not shown). 
Parameters as in Figure \ref{avr_vel}.
\label{mixing}}
\end{figure}

The dynamics of the system is well reflected in the information propagation
(Figure \ref{mixing}).
Both $\psi$ and $\mu$ have low values at small $s$ and have a maximum
value around $s_c$ (see also Ref. \cite{Sumpter_et_al_2008}). This result held for all density and noise parameter
values investigated. The curves of $\psi$ and $\mu$
were very similar, indicating that both are proper measures of information
propagation.

\begin{figure}[h!tbp]
{
\centerline{\includegraphics[width=0.5\columnwidth]{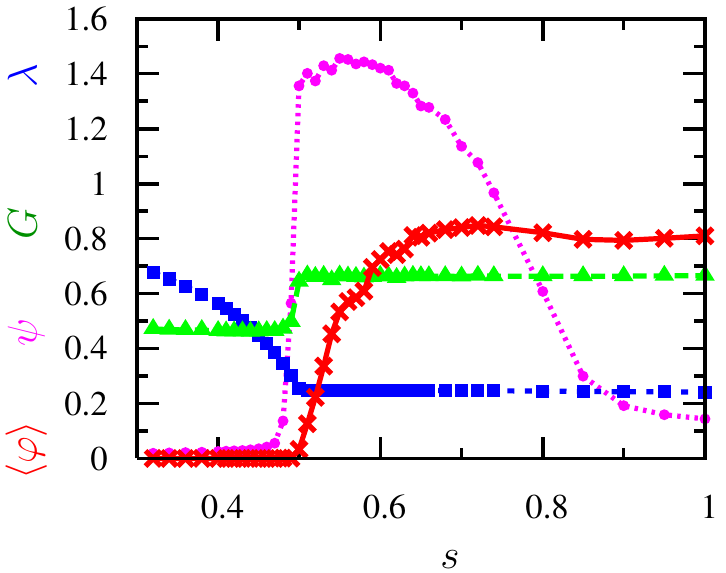}}}
\caption{The order parameter ($\varphi$, x marks), the Binder cumulant ($G$, triangles), 
the information exchange between particles ($\psi$, circles) and the average 
curvature of trajectories ($\lambda$, squares) as a function of the strategy 
variable $s$ in case of the AC-SPP with SVA.
Compared to the OVA in case of the SVA the critical
strategy value is at lower $s$, but the behavior of the $\varphi$, $G$ and $\lambda$ 
functions are very similar. The information exchange between particles has its maximum
value near the critical point as well, however, its variability as a function of $s$
is much stronger than for OVA. 
Parameters as in Figure \ref{avr_vel}.
\label{SVA}}
\end{figure}

In the original SPP model \cite{Vicsek_et_al_1995} (here, following the notation of
Huepe and Aldana, Ref \cite{Huepe_2008}, called as the Original Vicsek Algorithm 
(OVA)) the positions of the particles
at $t+\Delta t$ are depending on two previous time-steps, $t$ 
and $t-\Delta t$.
In the literature some authors have implemented the model in a slightly different way 
\cite{Gregoire_and_Chate_2004, Aldana_2003, Huepe_2004, Aldana_et_al_2007},
where the order of the position and the velocity update are changed.
Huepe and Aldana \cite{Huepe_2008} refer to this as the Standard Vicsek Algorithm (SVA)
and they report that the local density is different in the OVA and the SVA, 
while in both  
cases the average number of interacting neighbors is unreasonably high because
of the lack of a repulsive effect. 
By analysing the AC-SPP model with SVA-like updating rule (Figure \ref{SVA}), we find 
that the behavior of the order parameter, the Binder cumulant and the average 
curvature are very similar to those in the OVA, but the critical $s$ value is 
lower. The information exchange rate is, however, rather different and the maximum value of $\psi$ is 
much higher in case of the SVA. The maximum value of the information exchange
compared to the value at $s=1$ in case of the OVA is
$\psi^{OVA}_{max} / \psi^{OVA}_{s=1} = 1.4$, while in case of the SVA
$\psi^{SVA}_{max} / \psi^{SVA}_{s=1} = 10.1$.
At $s=1$, where the AC-SPP is congruous to the original SPP model, the rate of  
information exchange is similar in the OVA and the SVA.

\section{Conclusions}

In conclusion, we investigated the statistical properties of a
three-dimensional self-driven particle system (AC-SPP), designed to be an
improved model for the collective motion of living beings and possibly nonliving 
units (robots). The ordering of
particles exhibited a second-order phase transition as a function
of the control parameter corresponding to a behavioral strategy in
our case.

We found that the information exchange between particles was maximal at the
critical point. Due to the important role of information exchange in animal
societies, this might indicate that the critical point corresponds to an optimal
behavioral strategy. In a more general context, this result implies that biological
evolution may drive individual traits corresponding to critical values.
However, because of the individual based optimality requirement, this possibility needs
to be investigated within an evolutionary game theoretical framework.\\

{\bf ACKNOWLEDGMENTS.}
This work was supported by the EU FP6 Grant "Starflag" and 
Orsz\'agos Tudom\'anyos Kutat\'asi Alapprogramok (OTKA) 049674.


\begin{thebibliography}{99}

\bibitem{Inagaki_et_al_1976}
T. Inagaki, W. Sakamoto, and T. Kuroki,
Bull. Jap. Soc. Sci. Fish, \textbf{42}, 265 (1976).

\bibitem{Feare_book_1984}
C. J. Feare,
{\em The starlings}
(Oxford University Press, 1984).

\bibitem{Ben-Jacob_et_al_1995}
E. Ben-Jacob, I. Cohen, O. Shochet, A. Czir\'ok, and T. Vicsek,
Phys. Rev. Lett. \textbf{75}, 2899 (1995).

\bibitem{Rauch_et_al_1995}
E. M. Rauch, M. M. Millonas, and D. R. Chialvo,
Phys. Lett. A \textbf{207}, 185 (1995).

\bibitem{Shapiro_1995}
J. A. Shapiro,
BioEssays \textbf{17}, 597 (1995).

\bibitem{Parrish_1999}
J. K. Parrish and L. Edelstein-Keshet,
Science \textbf{284}, 99 (1999).

\bibitem{Szabo_et_al_2006}
B. Szabo, G. J. Szollosi, B.  Gonci, Z. Juranyi, D. Selmeczi, and T. Vicsek.
Phys. Rev. E \textbf{74}, 061908 (2006).

\bibitem{Wu_2006}
M. Wu, J. W. Roberts, S. Kim, D. L. Koch, and M.P. DeLisa
Applied and Enviromental Microbiology \textbf{72}, 4987-4994 (2006).

\bibitem{Parrish_Hammer_book_1997}
J. K. Parrish and W. M. Hammer,
{\em Animal Groups in Three Dimensions}
(Cambridge University Press, 1997).

\bibitem{Couzin_and_Krause_2003}
I. D. Couzin and J. Krause,
Adv. Study Behav. \textbf{32}, 1-75 (2003).

\bibitem{Vicsek_et_al_1995}
T. Vicsek, A. Czir\'ok, E. Ben-Jacob, I. Cohen, and O. Shochet,
Phys. Rev. Lett. \textbf{75}, 1226 (1995).

\bibitem{Reynolds_1987}
C. W. Reynolds,
Computer Graphics \textbf{21}, 25-33 (1987).

\bibitem{Czirok_et_al_1997}
A. Czir\'ok, H. E. Stanley, and T. Vicsek,
J. Phys. A  \textbf{30}, 1375 (1997).

\bibitem{Csahok_and_Vicsek_1995}
Z. Csah\'ok and T. Vicsek,
Phys. Rev. E \textbf{52}, 5297 (1995).

\bibitem{Toner_and_Tu_1995}
J. Toner and Y. Tu,
Phys. Rev. Lett. \textbf{75}, 4326 (1995).

\bibitem{Toner_and_Tu_1998}
J. Toner and Y. Tu,
Phys. Rev. E \textbf{58}, 4828 (1998).

\bibitem{Toner_et_al_1998}
J. Toner, Y. Tu, and M. Ulm, 
Phys. Rev. Lett. \textbf{80}, 4819 (1998).

\bibitem{Czirok_et_al_1999}
A. Czir\'ok, M. Vicsek, and T. Vicsek,
Physica A \textbf{264}, 299-304 (1999).

\bibitem{Gregoire_and_Chate_2001}
G. Gr\'egoire, H. Chat\'e, and Y. Tu,
Phys. Rev. E \textbf{64}, 011902 (2001).

\bibitem{Gregoire_and_Chate_2004}
G. Gr\'egoire and H. Chat\'e,
Phys. Rev. Lett. \textbf{92}, 025702 (2004).

\bibitem{Nagy_et_al_2007}
M. Nagy, I. Daruka, and T. Vicsek,
Physica A  \textbf{373}, 445-454 (2007).

\bibitem{Huepe_2008}
C. Huepe and M. Aldana,
Physica A. \textbf{387}, 2809-2822 (2008).  

\bibitem{Chate_et_al_2008}
H. Chat\'e, F. Ginelli, G. Gregoire, and F. Raynaud,
Phys. Rev. E \textbf{77}, 046113 (2008).

\bibitem{Ballerini_et_al_2008}
M. Ballerini, N. Cabibbo, R. Candelier, A. Cavagna, E. Cisbani, I. Giardina, V. Lecomte, A. Orlandi, G. Parisi, A. Procaccini, M. Viale, and V. Zdravkovic,
Proc. Natl. Acad. Sci. USA \textbf{105}, 1232-1237 (2008).

\bibitem{Theraulaz_et_al_2008a}
J. Gautrais, C. Jost, and G. Theraulaz,
Ann. Zool. Fennici, in press (2008).

\bibitem{Theraulaz_et_al_2008b}
J. Gautrais, C. Jost, M. Soria, A. Campo, S. Motch, R. Fournier, S. Blanco,
and G. Theraulaz,
J. Math. Biol., in press (2008).

\bibitem{Gonci_et_al_2008}
B. G\"onci, M. Nagy, and T. Vicsek,
EPJ Special Topics \textbf{157}, 53-59 (2008).

\bibitem{Aldana_2003}
M. Aldana and C. Huepe, 
J. Stat. Phys. \textbf{112}, 135-153 (2003). 

\bibitem{Huepe_2004}
C. Huepe and M. Aldana,
Phys. Rev. Lett. \textbf{92}, 168701 (2004).

\bibitem{Aldana_et_al_2007}
M. Aldana, V. Dossetti, C. Huepe, V.M. Kenkre, and H. Larralde,
Phys. Rev. Lett. \textbf{98}, 095702 (2007).

\bibitem{Binder_and_Herrmann_book_1997}
K. Binder and D. W. Herrmann,
{\em Monte Carlo Simulation in Statistical Physics: An Introduction}
(Springer, Berlin, 1997).

\bibitem{Sumpter_et_al_2008}
D. Sumpter, J. Buhl, D. Biro, and I. Couzin,
Theory Biosci. \textbf{127}(2), 177-186 (2008).

\end{thebibliography}
\end{document}